\documentclass[conference]{IEEEtran}
\usepackage{graphicx}
\usepackage{amsmath}
\usepackage{subfigure}
\usepackage{url}
\usepackage{color}

\usepackage{graphicx}
\usepackage{amssymb}
\usepackage{epstopdf}
\usepackage{amsmath}
\usepackage[ruled,vlined,linesnumbered]{algorithm2e}
\usepackage{geometry}
\newtheorem{theorem}{\textbf{Theorem}}

\newtheorem{definition}{\textit{Definition}}

\ifCLASSINFOpdf
\else
\fi
\IEEEoverridecommandlockouts
\begin{document}
%
\title{A Truthful Auction based Incentive Framework for Femtocell Access}

\author{\IEEEauthorblockN{Sha Hua$^*$, Xuejun Zhuo$^{\dag}$ and Shivendra S. Panwar$^*$
\IEEEauthorblockA{\\$^*$Polytechnic Institute of New York University, Brooklyn, NY, 11201, USA}
\IEEEauthorblockA{shua01@students.poly.edu; pliu@poly.edu; panwar@catt.poly.edu}
\IEEEauthorblockA{$^\dag$Department of Computer Science, Tsinghua University, China}
\IEEEauthorblockA{zhuoxj07@mails.tsinghua.edu.cn}
\IEEEauthorblockA{\\This manuscript is accepted to IEEE Wireless Communications and Networking Conference (WCNC), 2013.}
\thanks{This work is supported by NSF Grant IIP-1127960, the New York State Center for Advanced Technology in Telecommunications (CATT) and the NSF Wireless Internet Center for Advanced Technology (WICAT).}
}

}

%


\maketitle

\begin{abstract}
As cellular operators are suffering from a data explosion problem, and users are consequently experiencing poor data services, the introduction of femtocells offers a cost-effective way to mitigate this problem. Femtocells enable larger network capacity by increasing spatial reuse of the spectrum and shortening the distance to the users. Existing work has shown that open access femtocells, which allow unregistered macro users to connect, are efficient in reducing inter-cell interference and offloading traffic. However, a major obstacle constraining the potential capability of femtocells and open access is the lack of incentives for privately-owned femtocells to serve unregistered users. Hence in this paper, we propose a Vickrey-Clarke-Groves (VCG) auction based incentive framework for accessing such selfish femtocells. We consider two scenarios: One scenario involves a single macro user and another scenario has multiple macro users. We design auction schemes for both scenarios and show analytically that our schemes are truthful and have low computational complexity. Extensive simulations validate these properties and show huge performance improvement to the macro users.
\end{abstract}


%
\IEEEpeerreviewmaketitle

\section{Introduction}

The recent popularization and evolution of cellular networks provide smartphone users with ubiquitous Internet connectivity. Most of the users' data access activities, including web surfing, multimedia streaming and online gaming, are being ``mobilized''. This creates explosive traffic demand for cellular operators, which exceeds their network capacity, and hence adversely affects the users' experience~\cite{iPhoneOverloadATT}. To address this issue, it is desirable if the cellular network capacity can be increased, while a part of the traffic can be effectively offloaded.

The advent of femtocells~\cite{Chandrasekhar@CMag08} unveils a cost-effective way to mitigate such a data capacity crisis. Using the home user's broadband service, a femtocell is a light-weight base station overlaid on the existing cellular network, providing high-speed data access over a short range. Femtocells increase the network capacity by reducing the transmission distance and increasing spatial reuse of the spectrum. In addition, it is widely acknowledged that controlled \textit{open-access} femtocells, which allow both registered and unregistered users to access, can mitigate inter-cell interference and achieve traffic offloading~\cite{Choi@GLOBECOM08}. Note other offloading methods also exist, such as using WiFi hotspots~\cite{Xuejun@ICNP11} and ad-hoc networks~\cite{Hua@TMM11, Hua@Globecom09}.

However, the major obstacle lying between the potential capabilities and the wide adoption of open-access femtocells is the lack of incentives for these privately-owned femtocells to allow access to unregistered users. \textit{Why would a femtocell owner be willing to serve an unknown user who does not contribute to paying for this resource?} Our answer to this question is to use market mechanisms to incentivize these \textit{selfish} femtocells to provide a service for monetary returns. An auction~\cite{AuctionBook} is one of the most popular trading forms as it allows effective price discovery and efficient resource allocation. 

In this paper, we consider the inherent \textit{selfish} nature of the femtocells and focus on engineering the \textit{incentives} to promote the femtocells to \textit{truthfully} lease the access opportunity to the macro user equipments (MUEs) by means of an auction. We use MUE to denote the unregistered users close to the femtocells, who desire larger data rate.  Our framework guarantees the femtocell owners earn enough compensation by providing data service, and thus is a ``win-win'' solution to both parties. 

Only a few recent papers address the incentive issues of accessing femtocells~\cite{Yanjiao@INFOCOM12, Yanjiao@TWC12}, but they consider different aspects from those presented here. The authors in~\cite{Yanjiao@INFOCOM12} design an auction mechanism for femtocells to bid their area access permissions to the wireless service providers. Our work focuses on the trade in access time between femtocells and MUEs, and so is fundamentally different. In~\cite{Yanjiao@TWC12}, a refunding framework based on a Stackelberg Game is proposed between MUEs and femtocells. However, it is not a truthful mechanism so any agent is able to cheat. By contrast, our framework guarantees \textit{truthfulness}, which enables the participating agents to faithfully reveal their true valuations of the resource. 

Specifically, we consider two typical scenarios with multiple femtocells. In the scenario with a single MUE (SingleMUE for short), the MUE can aggregate the throughput from neighboring femtocells, and we design a \textit{multi-unit reverse auction} mechanism for femtocells to compete in selling their access times. In the scenario with multiple MUEs (MultiMUE in short), a \textit{multi-unit double auction} mechanism is proposed to find a best matching between MUEs and femtocells. None of these issues can be simply addressed by conventional auction theories~\cite{AuctionBook}. Inspired by VCG-auctions, in both scenarios, we aim to maximize the \textit{system efficiency} and show that this can be achieved with polynomial-time complexity. It indicates that our framework is both efficient and simple to implement. Moreover, we rigorously proved the \textit{truthfulness} of the mechanisms, which effectively prevents market manipulation. Our scheme is also \textit{individual rational}. Extensive simulations, based on real urban neighborhood topology, further show that the network performance can be noticeably improved. Both MUEs and femtocells achieve larger utilities by efficient resource allocation.

The remainder of the paper is organized as follows. Section II presents the preliminaries of auction theory. In Section III, we provide problem formulations of our framework, and algorithms for their solution. Then we give proofs of some properties. The performance evaluation is presented in Section IV. Finally, the paper is concluded in Section V.

\section{Preliminaries on Auction Theory}

In this section, we briefly overview the concept of auctions and some relevant terms and notations. 

In economics, an auction is a typical method to determine the value of a commodity that has a variable price. Most auctions are \textit{forward auctions} which involves a single seller and multiple buyers. In this paper, we use a \textit{reverse auction} in the SingleMUE scenario, which involve a single buyer (MUE) and multiple sellers (femtocells). The sellers compete for selling the commodities by submitting bids, then the buyer decides on its purchase. In addition, we use a \textit{double auction} in MultiMUE scenario, where multiple buyers and sellers are included. They submit bids and asks to the auctioneer, who decides the result. 
The notation is introduced below. \\
\textbf{bid ($b_i$)}: the valuation of the resource submitted by bidder $i$, which is not necessarily true. An ask ($a_i$) of a seller in a double auction is defined similarly. \\
\textbf{Private Value ($v_i$)}: the true valuation for the resource by bidder $i$. This value is only known by the bidder. \\
\textbf{Price ($p_i$)}: the price actually paid by the buyer $i$ (or paid to the seller $i$). \\
\textbf{Utility ($u_i$)}: the residual value of the resource. For buyer $i$, it is $u_i = v_i - p_i$, while for seller $i$, it is $u_i = p_i - v_i$. \\
\textbf{Individual Rationality}: An auction is individual rational if all buyers and sellers are guaranteed to obtain non-negative utility. It is a common requirement for auction designs. \\
\textbf{Truthfulness}: An auction is truthful if for every buyer/seller, submitting bid $b_i = v_i$ (or ask $a_i = v_i$) is a weakly dominant strategy, which maximizes $i$'s utility regardless of the strategies chosen by all other bidders. As the most critical property of auction scheme, truthfulness prevents market manipulation by revealing the true valuations of each participating agent and facilitates resource allocation.

In this work, we aim to design auction mechanisms to maximize the system efficiency and having 1) individual rationality, 2) truthfulness and 3) polynomial-time computability.

\section{Problem Formulation}

In this section, we give the detailed formulation and analysis of our auction incentive framework. For simplicity of presentation, we use access time as the basic commodity to trade. We assume the system is time-slotted and the auction is performed in rounds, with each round consisting of $T$ slots.

\subsection{Auction Mechanism Design for SingleMUE} \label{sec:SingleMUEProb}

\begin{figure}[htbp]
    \centerline{\includegraphics[scale=0.28]{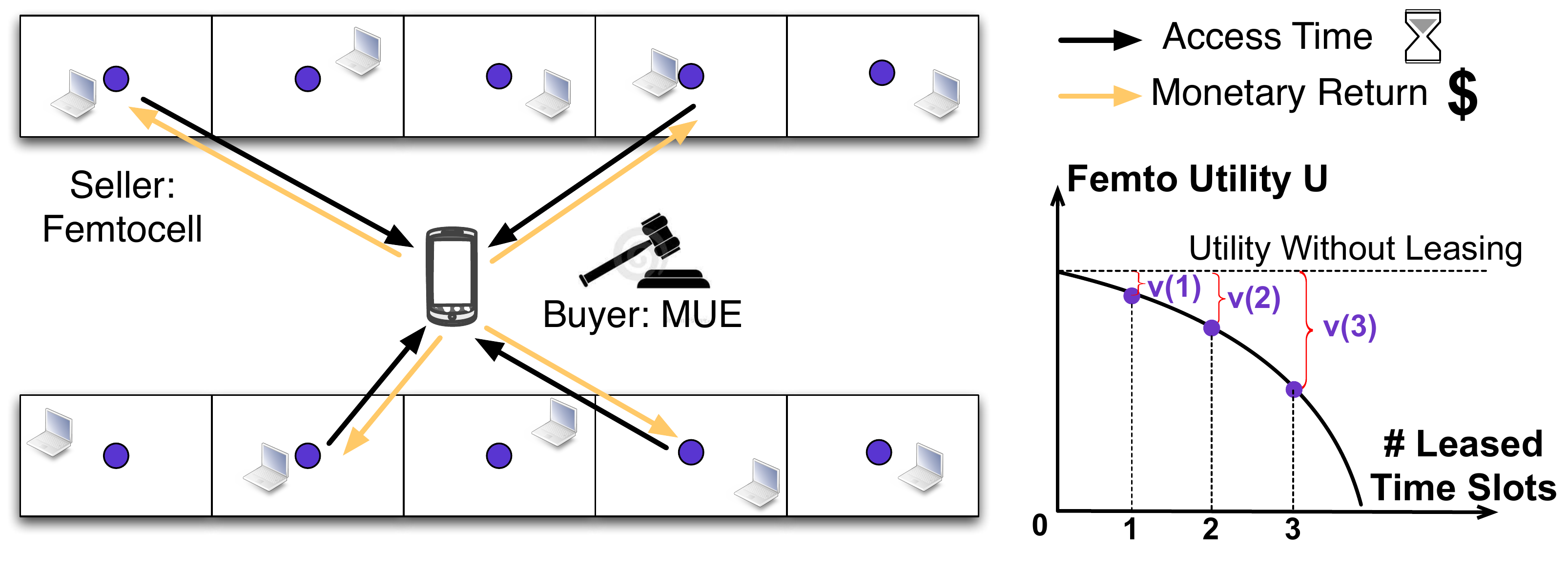}}
    \caption{
       SingleMUE: auction with one MUE and multiple femtocells.
    }
    \label{fig:onetomany}
    \vspace{-0.1in}
\end{figure}

The SingleMUE scenario involves one MUE and multiple neighboring femtocells, as shown in Figure~\ref{fig:onetomany}. This scenario happens when the operator prefers to set up an auction for each MUE. The femtocells are sellers selling the access time units to the MUE, who can aggregate the data from multiple femtocells to achieve a larger rate. We will design a multi-unit reverse auction framework with each time slot as a unit. The MUE is both the buyer and auctioneer, it receives the bids submitted by the femtocells and determines the result.

We use $\mathcal{I}$ to represent the set of femtocells, and $I = |\mathcal{I}|$ the total number. There is a limit on the maximum number of time slots that can be leased out, set by the owner for each femtocell $i$. We denote it as $N_{i}$. We further use $R_{i}$ to represent the data rate of the link between femtocell $i$ and MUE. It is measured by the MUE and does not change during an auction round. 

In our framework, each femtocell $i$ submits a bid vector:
\begin{equation}
\mathbf{b}_{i} = \{(0, 0), (1, b_i(1)), (2, b_i(2)), \ldots, (N_i, b_i(N_i))\},
\end{equation}
to the MUE, where $b_i(n), 1 \leq n \leq N_i$, is the bid for leasing $n$ time slots to the MUE. Additionally, as illustrated in the right part of Figure~\ref{fig:onetomany}, each femtocell $i$ has a private value for leasing $n$ time slots out, $v_i(n)$, which should be its own utility loss due to leasing. $v_i(n)$ may not necessarily be the same as the bid $b_i(n)$. However in a truthful auction, it can be proved that bidding the private value $b_i(n) = v_i(n)$ is always a weakly dominant strategy to the bidder. We will show this later.

Our auction mechanism design includes two parts, \textit{Winner Determination} and \textit{Pricing}. In the winner determination (W-D) part, the auctioneer, MUE, determines a set of femtocells as the winners according to the bid vectors they submitted. Following the VCG-based auction mechanisms, our winner determination aims to maximize the system efficiency, formulated as:
\begin{eqnarray}
\max_{n_i}: && U(\sum_{i \in \mathcal{I}} (\frac{n_i R_i}{T})) - \sum_{i \in \mathcal{I}} b_i(n_i) - U(R^{mac}) \label{eqn:2} \\
\mbox{subject to}: && \sum_{i \in \mathcal{I}}{n_i} \leq T, \label{eqn:3} \\
&& n_i \leq N_i, \qquad \forall i \in \mathcal{I}. \label{eqn:4}
\end{eqnarray}

In the above formulation, $U(\cdot)$ is the utility of the MUE as a function of its data rate. $n_i$ is the number of time slots leased to the MUE from femtocell $i$. $R^{mac}$ is MUE's data rate with the macro base station (BS). System efficiency is represented by the difference of the utility gain achieved by the MUE $U(\sum_{i \in \mathcal{I}} (n_i R_i)) - U(R^{mac})$, and the summation of the bids of all the sellers $\sum_{i \in \mathcal{I}} b_i(n_i)$, which is its typical definition~\cite{AuctionBook}. Constraints~(\ref{eqn:3}) and~(\ref{eqn:4}) limit the total leased time to not exceed $T$ and each $n_i$ to not exceed $N_i$.

We assume the MUE's utility function $U(\cdot)$ is concave in general, which reflects a wide rage of applications. For instance, video quality follows a log-like function with the received video rate. Therefore, for fixed $R_i$'s, $U(\sum_{i \in \mathcal{I}} (n_i R_i))$ is concave with $n_i$. In addition, in a truthful auction, $b_i(n_i) = v_i(n_i)$, and $v_i(n_i)$ is a convex function of $n_i$. To see this, consider there is one (or several) registered femtocell user(s) (FUE) associated with the femtocell $i$ with data rate $R_i^{\prime}$ and utility function $U^{\prime}(\cdot)$. The private value of leasing $n_i$ units of time will then be:
\begin{equation}
v_i(n_i) = U^{\prime}(R_i) - U^{\prime}(R_i^{\prime}(T - n_i)/T). \label{eqn:5}
\end{equation}
As we can see, increasing $n_i$ will reduce the utility of the registered user and then increase $v_i(n_i)$. It can be easily proved that when $U^{\prime}(\cdot)$ is concave, $v_i(n_i)$ is convex\footnote{In fact, besides the shown example, any proportional fair scheduling of FUEs and MUE makes $v_i(n_i)$ similar to the form of~(\ref{eqn:5}), and is convex.}. As a result, the objective shown in Equation~(\ref{eqn:2}) is concave with respect to $n_i$. 

Conventional winner determination, in the form of integer programming, results in high computational complexity and leads to untruthful auctions~\cite{Dobzinski@JAIR12}. However, when the objective is concave, satisfying the ``downward sloping property''~\cite{Dobzinski@JAIR12}, we can solve the above optimization problem in polynomial time. The procedures are given in Algorithm 1. It is a greedy algorithm that maximizes the marginal utility gain (Lines 5 to 8) in each iteration. It has a computational complexity of $O(\max\{T, \sum_{i}N_i\}|\mathcal{I}|)$ for each auction round.
\begin{algorithm}
\small{
$\Omega \leftarrow \mathcal{I}$\;
$n_i \leftarrow 0, \forall i \in \Omega$\;
\While{$\Omega \ne \emptyset$ and $T \ne 0$} {
	\ForEach{$i \in \Omega$} {
		$U_i^{gain} \leftarrow \mbox{ }\mbox{ }\mbox{ }\mbox{ }\mbox{ }\mbox{ } U(\sum_{j \in \mathcal{I}}(n_j R_j) + R_i)] - U(\sum_{j \in \mathcal{I}}(n_j R_j))$\;
		$U_i^{loss} \leftarrow \mbox{ }\mbox{ }\mbox{ }\mbox{ }\mbox{ }\mbox{ } \sum_{j \in \mathcal{I}, j \ne i}b_j(n_j) + b_i(n_i + 1) - \sum_{j \in \mathcal{I}}b_j(n_j)$\;
		$\Delta U_{i} \leftarrow U_i^{gain} - U_i^{loss}$\;
	}
	$i^{*} \leftarrow \arg\max_{i \in \Omega} \Delta U_i$\;
	$n_{i^{*}} \leftarrow n_{i^{*}} + 1$\;
	\If{$n_{i^{*}} == N_i$} {
		$\Omega \leftarrow \Omega \setminus i^{*}$\;
	} 
	$T \leftarrow T-1$\;
}
\If{$U(\sum_{i \in I} (n_i R_i)) < U(R^{mac})$} {
	$n_i \leftarrow 0, \forall i \in \mathcal{I}$\;
}
}
\label{alg:one}
\caption{W-D Algorithm for SingleMUE Auction}
\end{algorithm}

Next we focus on the pricing part. We have the following payment mechanism: 
\begin{definition} In SingleMUE auction, each winning femtocell $i$ receives a payment $p_i$ from the MUE as follows:
\begin{equation}
p_i = b_i(n_i^{*}) + (Q^{*} - Q^{*}_{-\mathbf{b}_i}), \label{eqn:6}
\end{equation}
where $Q^{*}$ and $Q^{*}_{-\mathbf{b}_i}$ are the optimal solutions of winner determination problem~(\ref{eqn:2})-(\ref{eqn:4}) with and without the bid vector $\mathbf{b}_i$ submitted by femtocell $i$. The utility achieved by it is
\begin{equation}
u_i = p_i - v_i(n_i^{*}).  \label{eqn:7}
\vspace{-0.03in}
\end{equation}
\end{definition}

Based on the above pricing rule, we then prove some important properties of our auction.

\begin{theorem} 
(\textit{Truthfulness}) For each femtocell $i$, setting its bid truthfully equal to its  private valuation, $\mathbf{b}_i = \mathbf{v}_i$, is a weakly dominant strategy. 
\end{theorem}
\begin{IEEEproof}
To prove the truthfulness, we compare two cases. One case is that the femtocell $i$ bids its true valuation as $\mathbf{b}_i = \mathbf{v}_i$. As shown from Equations~(\ref{eqn:6}) and~(\ref{eqn:7}), the resulting utility is $u_i = Q^{*} - Q^{*}_{-\mathbf{b}_i}$.

In the other case, the femtocell $i$ bids a false valuation $\mathbf{b}_i^{\prime} \ne \mathbf{v}_i$, aiming to improve its utility by cheating. As a result, the utility it can get becomes:
\begin{equation}
u_i^{\prime} = b_i^{\prime}(n_i^{*\prime}) + (Q^{*\prime} - Q^{*\prime}_{-\mathbf{b}_i^{\prime}}) - v_i(n_i^{*\prime}),
\end{equation}
where $Q^{*\prime}$ and $n_i^{*\prime}$ are the corresponding optimal solution of the bid vector set $\mathbf{B}^{\prime} = \{\mathbf{b}_1, \mathbf{b}_2, \ldots, \mathbf{b}_i^{\prime}, \ldots, \mathbf{b}_{I}\}$. Since $Q^{*}_{-\mathbf{b}_i} = Q^{*\prime}_{-\mathbf{b}_i^{\prime}}$, the loss of the utility can be computed as:
\begin{eqnarray}
\Delta u_{i} &=& u_{i} - u_{i}^{\prime} = Q^{*} - (Q^{*\prime} + b_i^{\prime}(n_i^{*\prime}) -  v_i(n_i^{*\prime})) \nonumber \\
&=& \big[U(\sum_{i \in \mathcal{I}} (n_i^{*} R_i /T)) - \sum_{i \in \mathcal{I}} b_i(n_i^{*})\big] \nonumber \\
&& - \big[U(\sum_{i \in \mathcal{I}} (n_i^{*\prime} R_i /T)) - \sum_{j \in \mathcal{I}, j \ne i} b_j(n_j^{*\prime}) - b_i(n_i^{*\prime})\big] \nonumber \\
&& - b_i^{\prime}(n_i^{*\prime}) + v_i(n_i^{*\prime}) \nonumber \\
&=& \big[U(\sum_{i \in \mathcal{I}} (n_i^{*} R_i /T)) - \sum_{i \in \mathcal{I}} b_i(n_i^{*})\big] \nonumber \\
&& - \big[U(\sum_{i \in \mathcal{I}} (n_i^{*\prime} R_i /T)) - \sum_{i \in \mathcal{I}} b_i(n_i^{*\prime})\big]
\end{eqnarray}
As $\{n_i^{*\prime}, i \in \mathcal{I}\}$ satisfies the constraints~(\ref{eqn:3}) and~(\ref{eqn:4}), it is then a feasible solution to the original problem shown in~(\ref{eqn:2}). Therefore, the optimal value achieved by $\{n_i^{*\prime}\}$ is always inferior to that achieved by $\{n_i^{*}\}$. Thus $\Delta u_{i} = u_{i} - u_{i}^{\prime} \ge 0$. It indicates that when the bid vectors submitted by other femtocells remain unchanged, femtocell $i$ cannot unilaterally increase its utility by submitting a bid vector $\mathbf{b}_i^{\prime}$ different from its private value. Therefore, bidding $\mathbf{v}_i$ is always a weakly dominant strategy, and we complete our proof.
\end{IEEEproof}
\begin{theorem}
(\textit{Individual Rationality}) The utility gained for each winning femtocell is a non-negative value.
\end{theorem}
\begin{IEEEproof}
As we have proved that our scheme is truthful, $b_i(n_i^{*}) = v_i(n_i^{*}), \forall i$, and the utility of a winning femtocell $i$ is $u_i = Q^{*} - Q^{*}_{-\mathbf{b}_i}$. Note that $Q^{*}_{-\mathbf{b}_i}$ can be viewed as the optimal solution when femtocell $i$ submits the bid vector as $\mathbf{b}_i = \{(0, 0), (1, \infty), \ldots, (N_i, \infty)\}$ and the bid vectors from other femtocells remain unchanged. It is equivalent to fixing $n_i = 0$ in the solution. The resulting solution set with $n_i = 0$ is a subset of the solution space to resolve $Q^{*}$. Therefore, the value of $Q^{*}_{-\mathbf{b}_i}$ is always inferior to $Q^{*}$, and $u_i \geq 0$.
\end{IEEEproof}

\subsection{Auction Mechanism Design for MutliMUE}

\begin{figure}[htbp]
    \centerline{\includegraphics[scale=0.3]{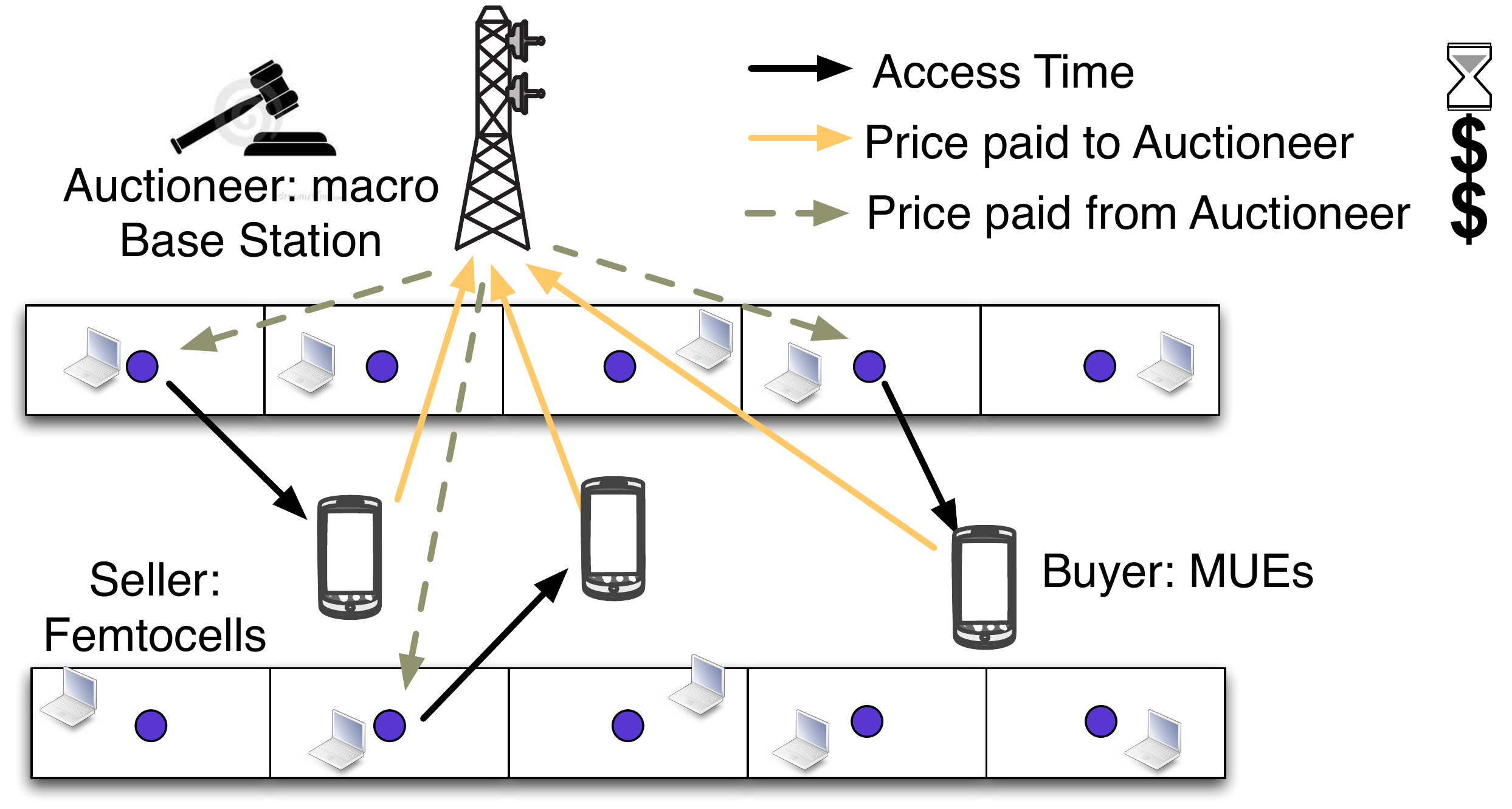}}
    \caption{
       MultiMUE: auction with multiple MUEs and femtocells.
    }
    \label{fig:manytomany}
    \vspace{-0.1in}
\end{figure}

Another scenario we consider, MultiMUE, involves multiple MUEs and multiple femtocells coexisting in one area, as shown in Figure~\ref{fig:manytomany}. This happens when the operator prefers to set up an auction covering multiple MUEs. Femtocells are still the resource holders selling their access time slots to the MUEs. The MUEs are buyers. We assume the macrocell covering the area is the auctioneer, who receives the bids/asks from all the agents and then performs winner determination and pricing. This characterizes a scenario similar to the exchange market and can be analyzed using multi-unit double auction.

We use $\mathcal{I}$ and $\mathcal{J}$ respectively to represent the set of MUEs and femtocells. In a double auction, each MUE $i \in \mathcal{I}$ submits a set of bid vectors $\{\mathbf{b}_{ij}\}, j \in \mathcal{J}$ to the auctioneer, where 
\begin{equation}
\mathbf{b}_{ij} = \{(0, 0), (1, b_{ij}(1)), (2, b_{ij}(2)), \ldots, (T, b_{ij}(T))\}.
\end{equation} 
Each term $b_{ij}(k)$ stands for the bid for buying $k$ time slots of access time from femtocell $j$. As femtocells do not distinguish users, each femtocell submits an ask vector
\begin{equation}
\mathbf{a}_{i} = \{(0, 0), (1, a_i(1)), (2, a_i(2)), \ldots, (N_i, a_i(N_i))\},
\end{equation}
with $a_i(k)$ representing the ask for leasing $k$ time units out. As in Section~\ref{sec:SingleMUEProb}, we aim to find a truthful double auction mechanism that drives both the buyers and sellers to submit their bids and asks equivalent to their private values. Specifically for each MUE $i \in \mathcal{I}$, the true valuation $v_{ij}^{b}(k) = U(kR_{ij}) - U(R_i^{mac})$, where $R_{ij}$ is the data rate for accessing femtocell $j$ in a unit time slot. For each femtocell $j \in \mathcal{J}$, $v_j^{s}(k)$ should be the utility loss resulting from leasing $k$ time slots. 

To simplify the modeling, we restrict to the case that one MUE can only buy time slots from one femtocell, and one femtocell can only sell time slots to one MUE. We use $n_{ij}$ to denote the number of time slots bought from femtocell $j$ to MUE $i$. Then we can maximize the overall system efficiency as follows: 
\begin{eqnarray}
\max_{n_{ij}}: && \sum_{i \in \mathcal{I}} \sum_{j \in \mathcal{J}} b_{ij}(n_{ij}) - 
\sum_{j \in \mathcal{J}} \sum_{i \in \mathcal{I}} a_j (n_{ij}) \label{eqn:15} \\
&& \nonumber \\
\mbox{subject to}: && 0 \leq n_{ij} \leq N_j, \forall j \in \mathcal{J}, \label{eqn:16} \\
&& \sum_{i}n_{ij} = \max_{i}\{n_{ij}\}, \forall j \in \mathcal{J}, \label{eqn:17} \\
&& \sum_{j}n_{ij} = \max_{j}\{n_{ij}\}, \forall i \in \mathcal{I}, \label{eqn:18}
\end{eqnarray}
where~(\ref{eqn:15}) follows the typical definition. Constraints~(\ref{eqn:17}) and~(\ref{eqn:18}) reflect the one-to-one mapping relationship. In the winner determination phase, we aim to solve the above problem in polynomial time. Fortunately, problem~(\ref{eqn:15})-(\ref{eqn:18}) can be converted into a max-weight bipartite matching problem.
\begin{figure}[htbp]
    \centerline{\includegraphics[scale=0.3]{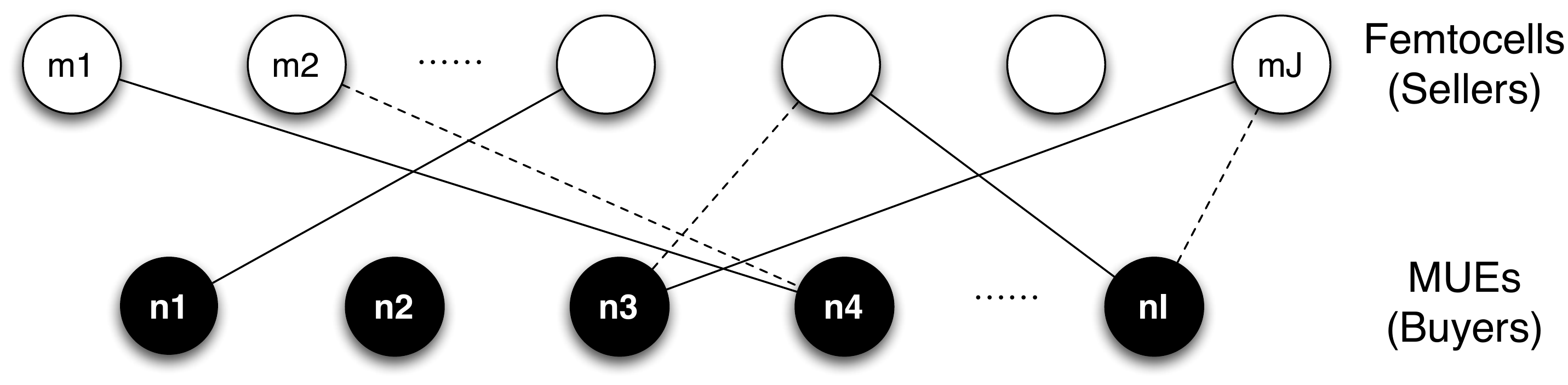}}
    \caption{
       Bipartite matching graph for femtocells and MUEs.
    }
    \label{fig:bipartite}
    \vspace{-0.2in}
\end{figure}

The bipartite graph is constructed as shown in Figure~\ref{fig:bipartite}. Each femtocell $j \in \mathcal{J}$ is represented as a vertex $m_j$ in the upper part of the graph and each MUE $i \in \mathcal{I}$ is represented as a vertex $n_i$ in the lower part. The weight on each edge connecting the femtocell $i$ and MUE $j$ is set to $w_{ij} = \max_{n_{ij}, n_{ij} \leq N_j}(b_{ij}(n_{ij}) - a_j(n_{ij}))$. Once the bipartite graph is formed, a matching that maximizes the sum weight can be found using Hungarian algorithm~\cite{Christos@BOOK07}, with complexity $O(I + J)$. We summarize the procedure in Algorithm 2.
\begin{algorithm}
\small{
Construct the bipartite graph\;
$w_{ij} \leftarrow \max_{n_{ij}, n_{ij} \leq N_j}(b_{ij}(n_{ij}) - a_j(n_{ij}))$\;
Find the optimal matching $\Phi$ using Hungarian Method. Nodes selected in $\Phi$ are the winners\;
If $i$ and $j$ are a pair in $\Phi$, $i$ will lease $n_{ij}^{*} = \arg\max_{n_{ij}, n_{ij} \leq N_j}(b_{ij}(n_{ij}) - a_j(n_{ij}))$ units from $j$\;
}
\label{alg:two}
\caption{W-D Algorithm for MultiMUE Auction}
\end{algorithm}

Then we have the following pricing mechanism:
\begin{definition}
In our auction mechanism design, each winning MUE $i \in \mathcal{I}$ pays the auctioneer:
\begin{equation}
p_i^{b} = b_{i, \sigma(i)}(n_{i, \sigma(i)}^{*}) - (Q^{*} - Q_{-\mathbf{b}_i}^{*})
\end{equation}
Each winning femtocell $j \in \mathcal{J}$ receives the payment from auctioneer with:
\begin{equation}
p_{j}^{s} = a_{\sigma(j), j}(n_{\sigma(j), j}^{*}) + (Q^{*} - Q_{-\mathbf{a}_j}^{*})
\end{equation}
\end{definition}
where $\sigma(\cdot)$ stands for the partner determined by W-D algorithm. Other terms remain the same meaning as in~(\ref{eqn:6}).
 
Next, we prove the properties of our double auction mechanism. It has been proved that no double auction can achieve efficiency, budget balance and truthfulness at the same time, even putting individual rationality aside~\cite{Myerson@JET83}. In MultiMUE, we aim to have efficiency, truthfulness and individual rationality. The macrocell BS, being the auctioneer, can tolerate a certain level of budget unbalance as an incentive to motivate users for traffic offloading. This concept has also been shown in~\cite{Xuejun@ICNP11, Xuejun@TMC13}. Further, a reserve price $P_{res}$ can be set by the auctioneer, and when its profit is below $P_{res}$, it can terminate the auction. Due to space limitation, we just show proofs for the MUE side. The femtocell side are similar due to symmetry. 
\begin{theorem} 
(\textit{Truthfulness}) For each MUE $i$, setting its bid truthfully as its  private value, is a weakly dominant strategy. 
\end{theorem}
\begin{IEEEproof}
Similar to the proof of Theorem 1, we compare two cases. In case one, MUE $i$ bids its private value $\mathbf{b}_i = \mathbf{v}_i$ and achieves the utility $u_i^b = Q^{*} - Q_{-\mathbf{b}_i}^{*}$.

In case two, the MUE will bid a false valuation $\mathbf{b}_i^{\prime} \ne \mathbf{v}_i$. Then the utility it can get becomes:
\begin{equation}
u_i^{b\prime} = v_{i, \sigma(i)}(n_{i, \sigma(i)}^{*\prime}) - b_{i, \sigma(i)}^{\prime}(n_{i, \sigma(i)}^{*\prime}) + (Q^{*\prime} - Q_{-\mathbf{b}_i^{\prime}}^{*\prime})
\end{equation}

As the bid/ask vectors submitted by others remain unchanged, we have $Q_{-\mathbf{b}_i^{\prime}}^{*\prime} = Q_{-\mathbf{b}_i}^{*}$. The changes of the utility can be computed as:
\begin{eqnarray}
&& \Delta u_i^b = u_i^b - u_i^{b\prime} \nonumber \\
&=& Q^{*} - \big[(Q^{*\prime} - b_{i, \sigma(i)}^{\prime}(n_{i, \sigma(i)}^{*\prime}) + v_{i, \sigma(i)}(n_{i, \sigma(i)}^{*\prime})\big] \nonumber \\
&=& Q^{*} - \big[(\sum_{j \ne i, j \in \mathcal{I}}b_{j, \sigma(j)}(n_{j, \sigma(j)}^{*\prime}) + v_{i, \sigma(i)}(n_{i, \sigma(i)}^{*\prime}) \nonumber \\
&& - \sum_{k \in \mathcal{J}}a_{k}(n_{\sigma(k), k}^{*\prime})\big] \nonumber \\
&=& Q^{*} - \big[\sum_{j \in \mathcal{I}}b_{j, \sigma(j)}(n_{j, \sigma(j)}^{*\prime}) - \sum_{k \in \mathcal{J}} a_{k}(n_{\sigma(k), k}^{*\prime})\big].
\end{eqnarray}

As the set of solutions $\{n_{ij}^{*\prime}, i \in \mathcal{I}\}$ satisfies the constraints~(\ref{eqn:16})-(\ref{eqn:18}), it is then a feasible solution for the problem shown in~(\ref{eqn:15}). Therefore, the value achieved by $\{n_{ij}^{*\prime}, i \in \mathcal{I}\}$ is always inferior to that achieved by the optimal solution of~(\ref{eqn:15}). Therefore $\Delta u_{i}^{b} \ge 0$. It indicates that when the bid vectors submitted by other MUEs and femtocells remain unchanged, MUE $i$ cannot unilaterally increase its utility by submitting a bid vector different from its private value. Bidding $\mathbf{v}_i^{b}$ is always a weakly dominant strategy.
\end{IEEEproof}
\begin{theorem}
(\textit{Individual Rationality}) The utility gained for each winning MUE is a non-negative value.
\end{theorem}
\begin{IEEEproof}
As our double auction mechanism is truthful, $b_{i, \sigma(i)}(n_{i, \sigma(i)}^{*}) = v_{i, \sigma(i)}(n_{i, \sigma(i)}^{*})$. The utility that MUE $i$ achieves is then:
\begin{equation}
u_i^{b} = v_{i, \sigma(i)}(n_{i, \sigma(i)}^{*}) - p_i^{b} =  Q^{*} - Q_{-\mathbf{b}_i}^{*} \label{eqn:20}
\end{equation}
~(\ref{eqn:20}) has the same form as what is in the proofs for Theorem 2. With same steps, we know $u_i^{b} \ge 0$ always holds.
\end{IEEEproof}

%

\section{Performance Evaluation}
In this section, we present extensive experiments to evaluate the performance of the auction schemes we proposed. 

\subsection{Experimental Setup}
\begin{figure}[htbp]
    \centerline{\includegraphics[scale=0.44]{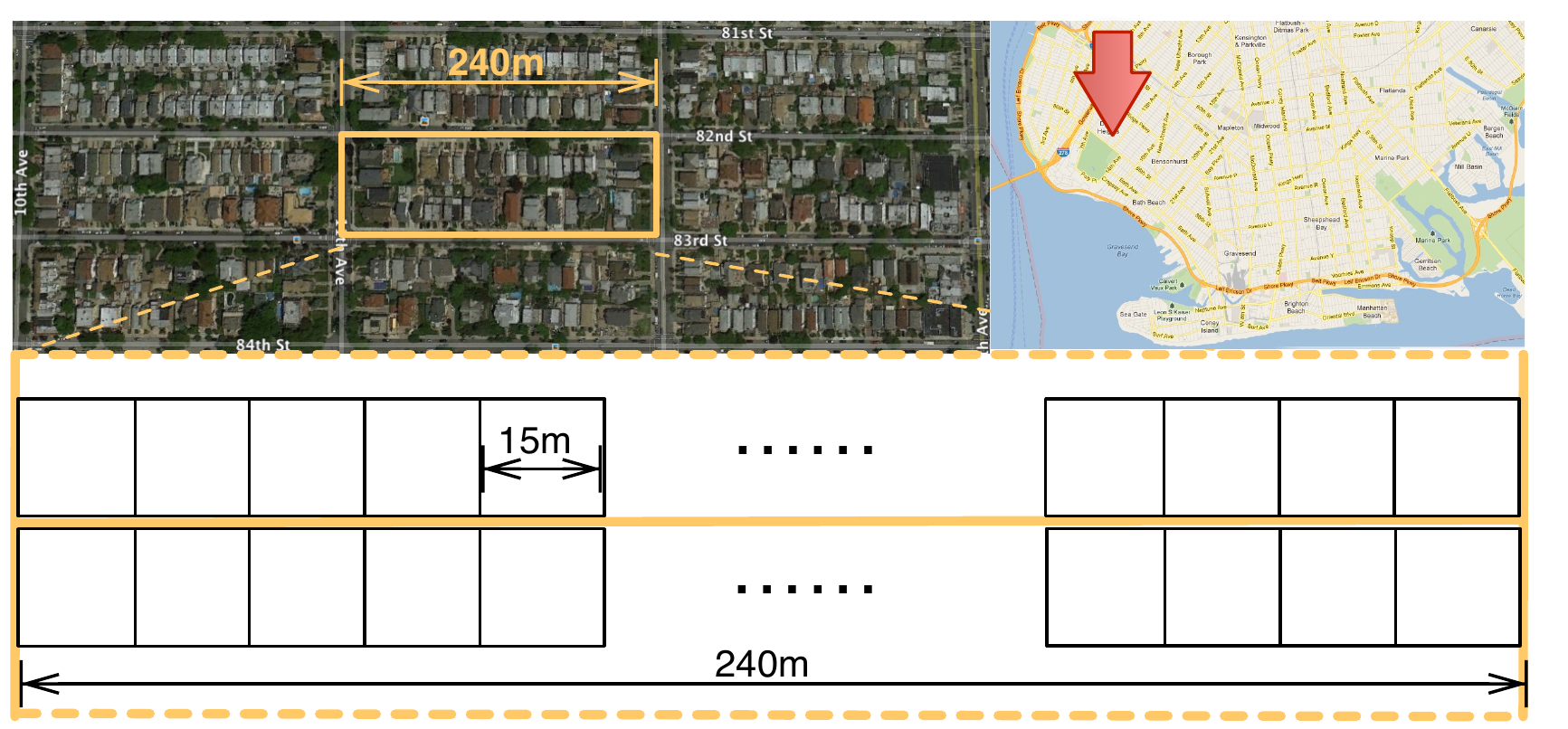}}
    \caption{
       Area map of a typical neighborhood in Brooklyn, NY.
    }
    \label{fig:brooklyn}
\end{figure}

In our simulations, we consider a typical urban neighborhood in Brooklyn, NY, USA, as shown in Figure~\ref{fig:brooklyn}. We pick one road segment for simulation. The length of the road is 240m. The apartments are squares, distributed along the two sides of the road, of size 15m x 15m each. We use a parameter, \textit{FemtoDensity}, to denote the probability that an apartment has a femtocell. Three FUEs exist in each apartment that has a femtocell. Femtocells and FUEs are randomly placed in the apartment. The macro BS is located at 300 meters north of the center point of the figure, with 500 active macro users.

We follow the IEEE 802.16m evaluation methodology document~\cite{WiMAXEva} for the channel model of macro BS$\leftrightarrow$MUE link. For the femto$\leftrightarrow$UE links, the standard industry model for femtocell evaluation~\cite{FemtoForum} is used. For all the channels, the fast fading component is modeled as Rayleigh fading with $\sigma = 1$. Shadow fading is modeled as a log-normal random variable with a standard deviation of 8 dB. The main system parameters and path loss at distance $d$ are summarized in Table I. We assume the auction round $T = 100$ time slots. Sigmoid function is used to represent the utility of each UE: $U(R) = 1 - e^{-a \frac{R}{R_{dem}}}$, where $a$ is the satisfactory factor fixed to 1, $R$ is the received data rate in the current round, and $R_{dem}$ is the traffic demand. $R_{dem}$ of the MUEs are fixed to 4 Mb/s, while those of the FUEs are uniformly distributed within the range $[0, R_{dem}^{max}]$.
\begin{table}[!t]
\caption{System Parameters}
\label{table:paras}
\centering
\begin{tabular}{| l | c |}
\hline
BS Transmit Power & 46 dBm \\
\hline
Noise power spectrum density $N_0$ & -174 dBm \\
\hline
Macro channel bandwidth & 10 MHz \\
\hline
Macro BS antenna gain & 7 dB \\
\hline
Femto BS Transmit Power & 0 dBm \\
\hline
Femto channel bandwidth & 5 MHz \\
\hline
Frequency reuse factor of femtocells & 6 \\
\hline
UE noise figure & 4 dB \\
\hline
Wall loss & 10 dB \\
\end{tabular}
\begin{tabular}{| c | c |}
\hline
Path loss (macrocell) & $L^{macro}(d)_{dB} = 17.39 + 3.76\log_{10}d$ \\
\hline
Path loss (femtocell) & $L^{femto}(d)_{dB} = 38.46 + 20\log_{10}d + 0.7d$ \\
\hline
\end{tabular}
\vspace{-0.2in}
\end{table}
\subsection{Truthfulness of the Auction Schemes}
\begin{figure}[htbp]
    \centerline{\includegraphics[scale=0.48]{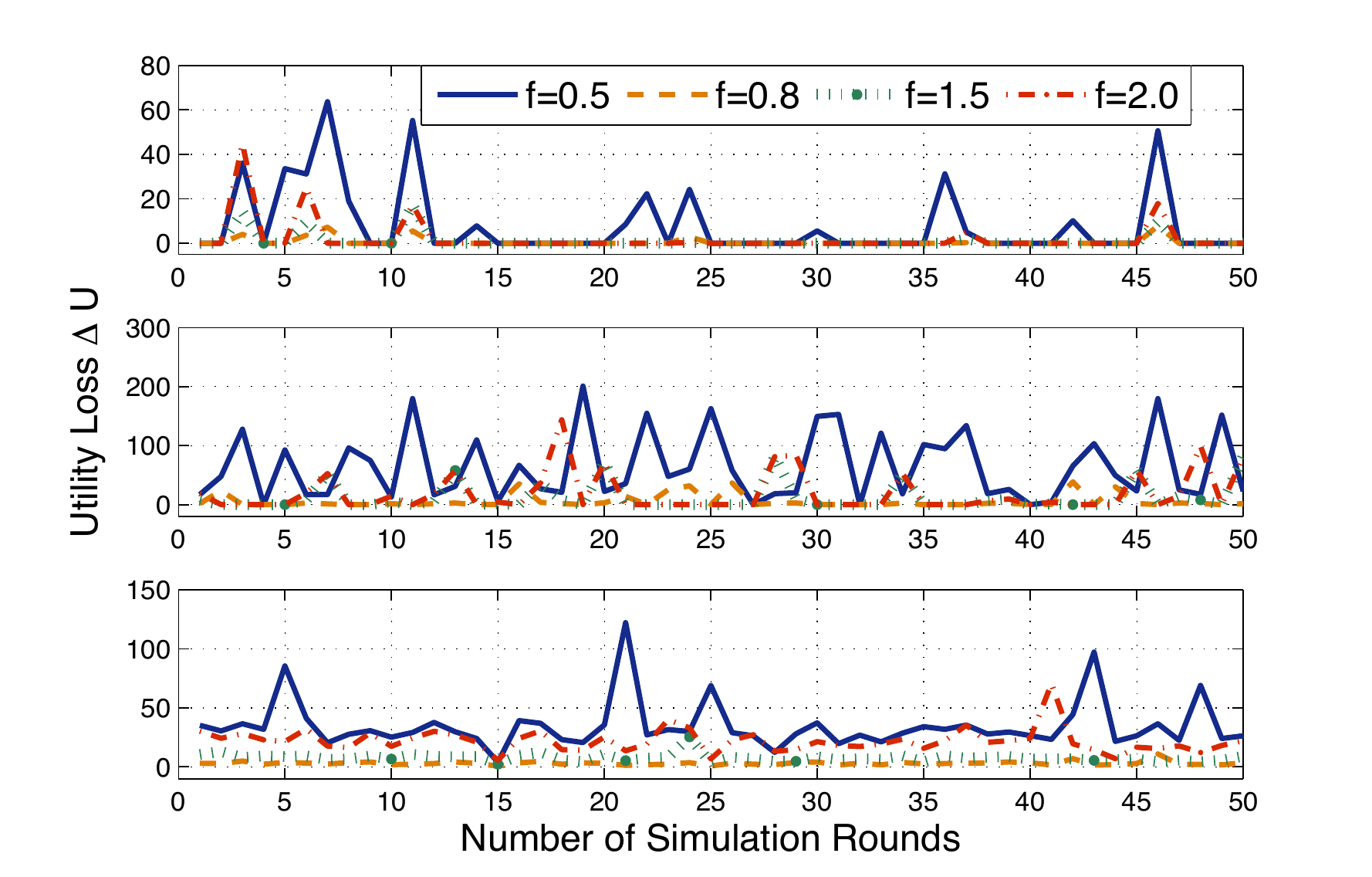}}
    \caption{
       Utility loss $\Delta U = U - U^\prime$ by submitting untruthful bids/asks: (a) A femtocell in SingleMUE scenario; (b) A femtocell in MultiMUE scenario; (c) An MUE in MultiMUE scenario.
    }
    \label{fig:truthful}
    \vspace{-0.2in}
\end{figure}

Here we validate the truthfulness of our proposed auction schemes. For the SingleMUE scenario, we choose a reference femtocell and compare the utilities it can get by bidding truthfully $\mathbf{b} = \mathbf{v}$ and untruthfully $\mathbf{b^\prime} = \mathbf{v} \cdot f$, where $f$ is a scaling factor. Similarly for the MultiMUE scenario, we choose a reference femtocell and an MUE, and manipulate their ask and bid vector to $\mathbf{a^\prime} = \mathbf{v} \cdot f$ and $\mathbf{b^\prime} = \mathbf{v} \cdot f$, respectively, in two tests. \textit{FemtoDensity} is set to 1 and $R_{dem}^{max}$ is set to 6 Mb/s. The values of the utility loss $\Delta U = U - U^\prime$ caused by bidding untruthfully in 50 auction rounds are shown in Figure~\ref{fig:truthful}. It can be seen from all three figures that when $f = \{0.5, 0.8, 1.5, 2\}$, $\Delta U$ is always a non-negative value. Although we cannot test all the sets of untruthful bids/asks, as they form an infinite space, Figure~\ref{fig:truthful} gives an indication that any agent cannot unilaterally improve its utility by submitting fake bids/asks.
 \begin{figure*}[htbp]
    \centering
    \subfigure[]
    {
        \includegraphics[scale=0.4]{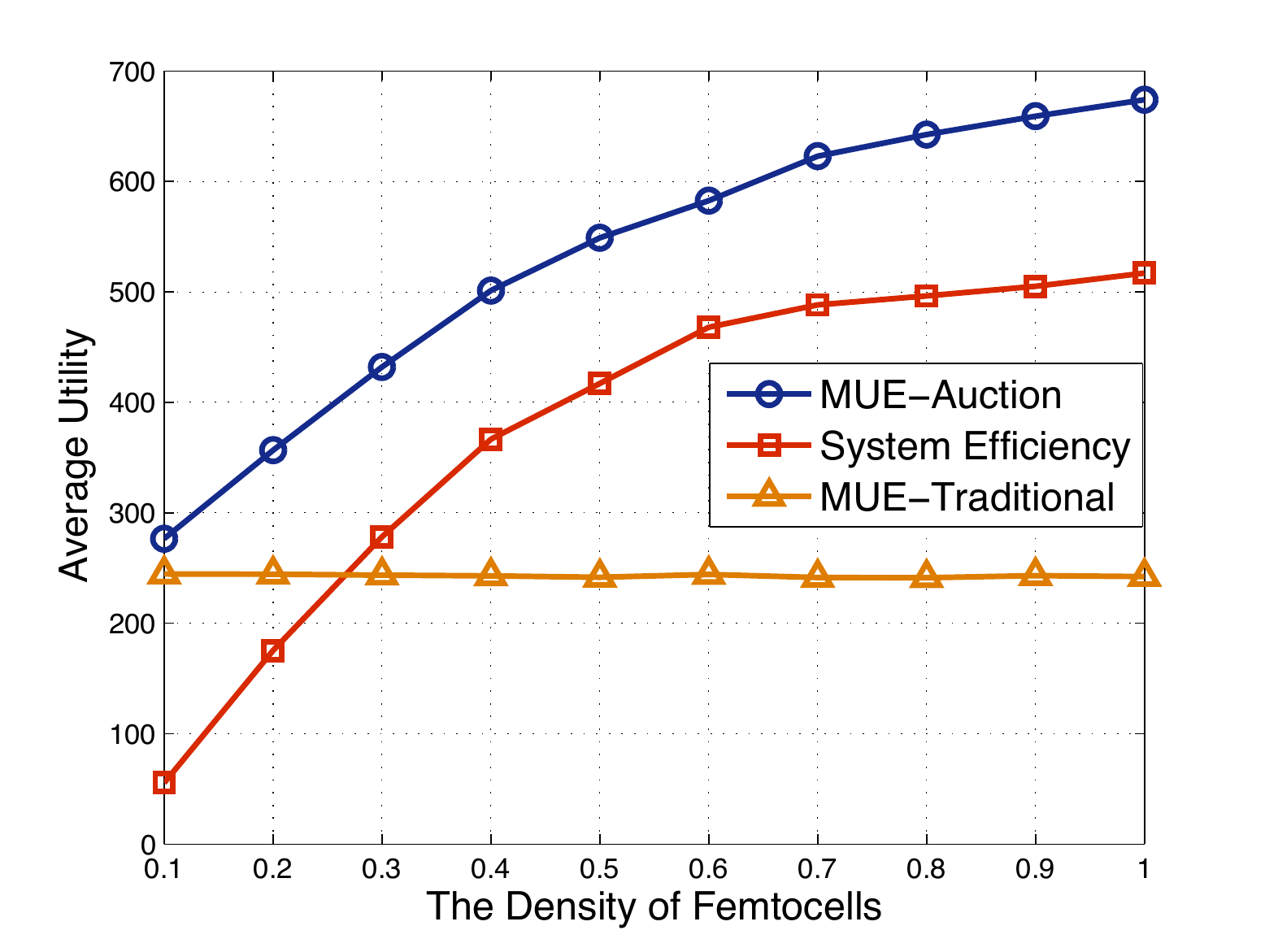}
        \label{fig:SingleMUEFemtoDensity}
    }
    \hspace{-0.35in}
    \subfigure[]
    {
        \includegraphics[scale=0.4]{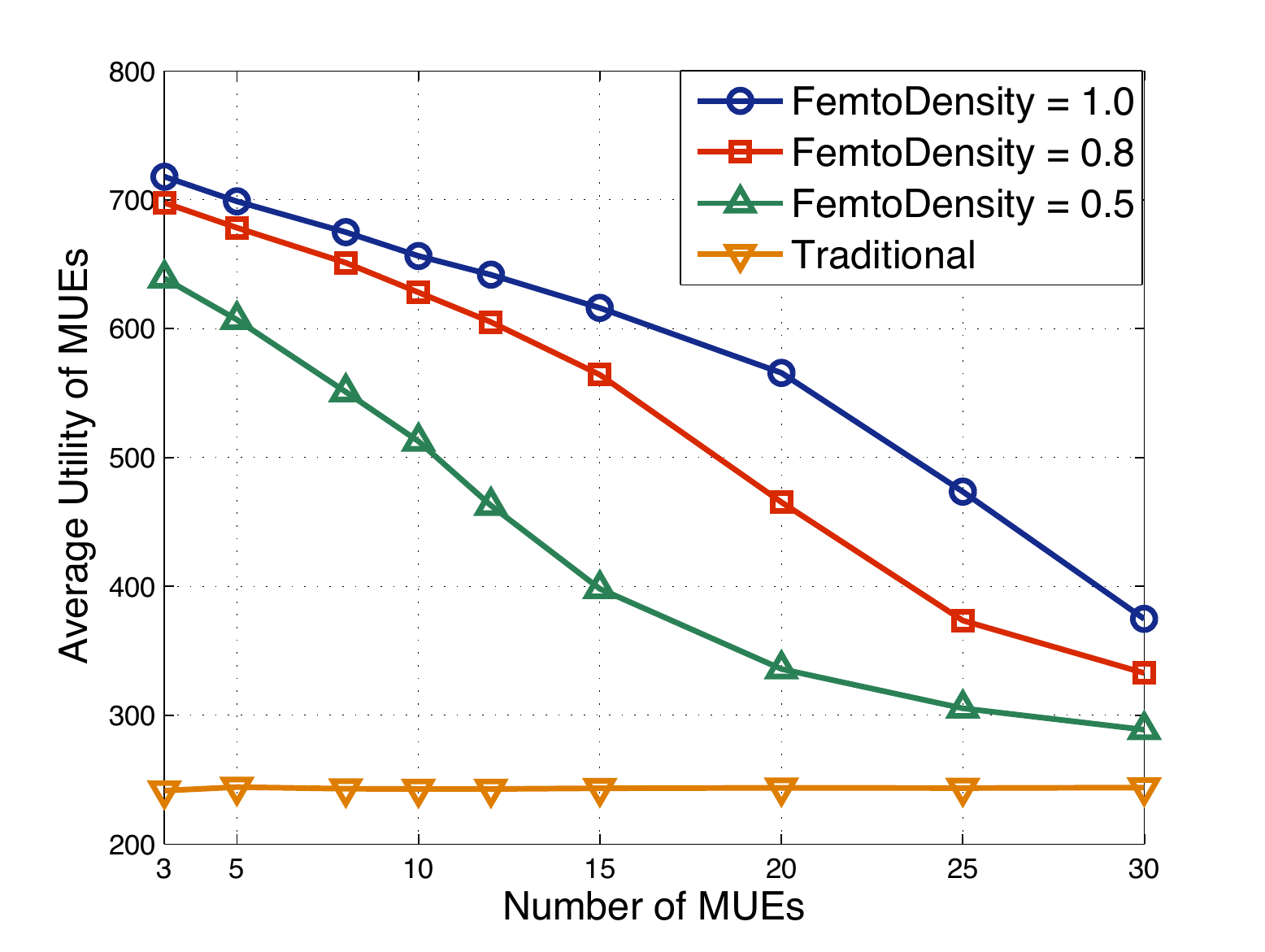}
        \label{fig:MultiMUEnumMUE}
    }
    \hspace{-0.35in}
    \subfigure[]
    {
        \includegraphics[scale=0.4]{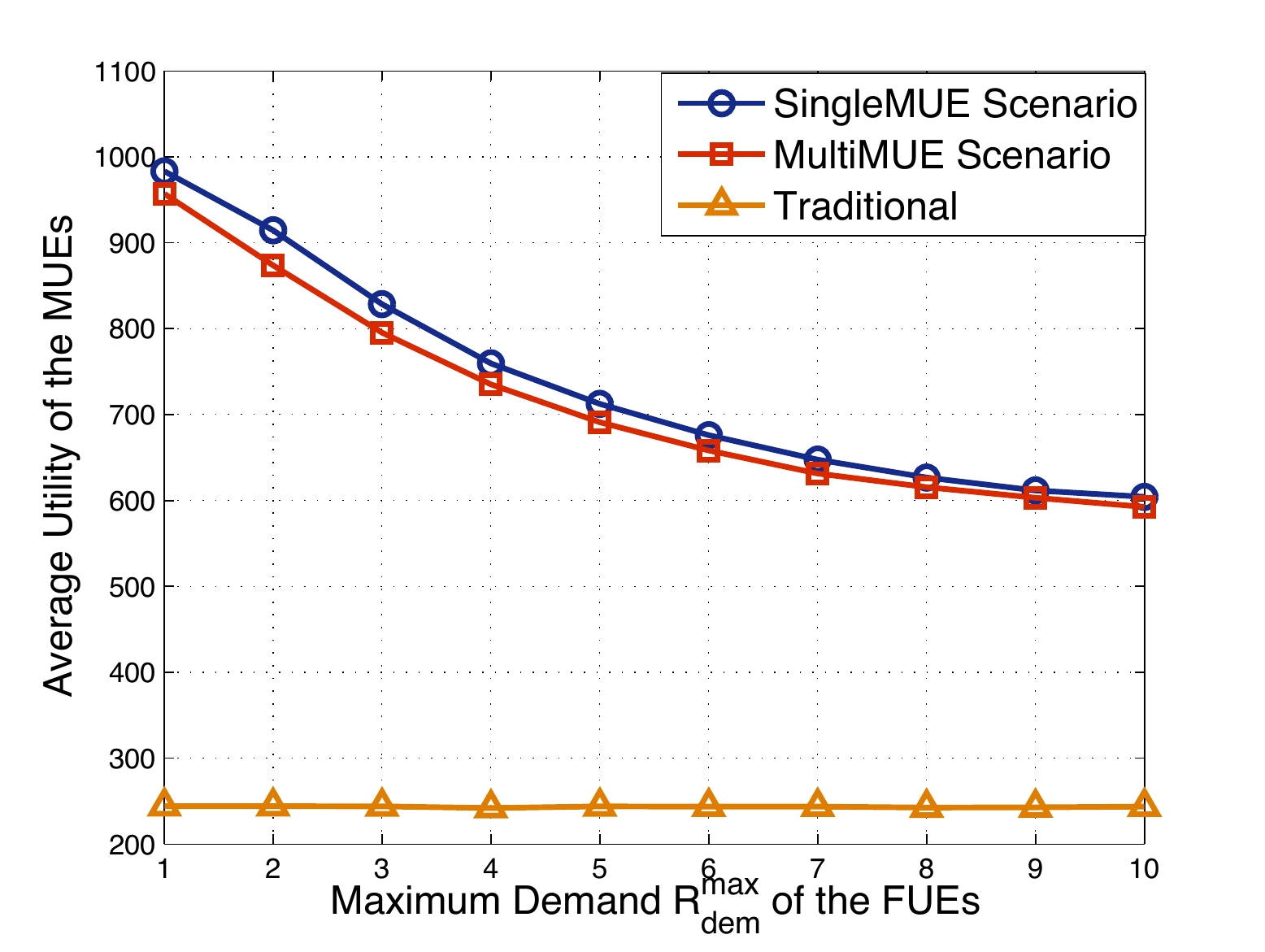}
        \label{fig:MUEmaxDemand}
    }
    \caption{System performance: (a) the impact of femtocell density in SingleMUE. (b) the impact of number of MUEs in MultiMUE (c) the impact of maximum demand of FUEs.}
    \label{fig:performance}
    \vspace{-0.1in}
\end{figure*}
\subsection{System Performance}
Next we evaluate the system performance of the two scenarios under the influence of several key parameters. All the points in the figures are the averages of 500 auction rounds.

First, for the SingleMUE scenario, we fix \textit{MaxDemand} $R_{dem}^{max}$ of the FUEs to 6 Mb/s and increase \textit{FemtoDensity} from 0.1 to 1.0. We can see from Figure~\ref{fig:SingleMUEFemtoDensity} that when the density of femtocells is close to 0, system efficiency is also close to 0, as few transactions happen. As the \textit{FemtoDensity} increases, the average utility of the MUEs increases. This is because there are more potential sellers providing better commodities. Our auction scheme significantly improves the utility of the MUE, by about 127\% when \textit{FemtoDensity = 0.5} and 166\% when \textit{FemtoDensity = 0.8}. Next, for MultiMUE scenario, Figure~\ref{fig:MultiMUEnumMUE} shows the impact of the number of MUEs. We can clearly see from the figure that as the number of MUEs increases, the average utility of MUEs drops. This is because the competition among MUEs forces them to increase the price and lowers each one's share of the femtocells, which follows basic economic principles. Also similar to Figure~\ref{fig:SingleMUEFemtoDensity}, the average utility of the MUEs increases when \textit{FemtoDensity} increases.

Next, we examine the impact of traffic demand $R_{dem}$ of FUEs. We fix \textit{FemtoDensit} to 1 and adjust $R_{dem}^{max}$ in the range of [1, 10] in the simulations. We deploy 10 MUEs in MultiMUE scenario. As shown in Figure~\ref{fig:MUEmaxDemand}, in both SingleMUE and MultiMUE scenarios, as $R_{dem}^{max}$ increases, the average utility of the MUEs drops. The reason is that when the traffic demand of FUEs gets higher, a femtocell's private value on leasing one unit access time will increase, resulting in higher price. However, we can still see our framework leads to 150\% improvement even when $R_{dem}^{max} = 10$ Mb/s, which is significant. Note all the results are based on the truthfulness of the schemes.

\section{Conclusion}

In this paper, we proposed an incentive framework to motivate femtocells to open their access to unregistered MUEs, which help increase network capacity and offload traffic. We carefully designed the VCG-based auction mechanisms to allocate access times and rigorously proved that all the participating agents can truthfully cooperate. Simulation results demonstrate that the performance of MUEs can be significantly improved, with system efficiency maximized in auctions. In the future, we plan to study the incentive issue of femtocells controlled by different operators, as in~\cite{Sha@ICC12}.

\bibliographystyle{IEEEtran}
\bibliography{pricingbib}

\begin{thebibliography}{10}
\providecommand{\url}[1]{#1}
\csname url@samestyle\endcsname
\providecommand{\newblock}{\relax}
\providecommand{\bibinfo}[2]{#2}
\providecommand{\BIBentrySTDinterwordspacing}{\spaceskip=0pt\relax}
\providecommand{\BIBentryALTinterwordstretchfactor}{4}
\providecommand{\BIBentryALTinterwordspacing}{\spaceskip=\fontdimen2\font plus
\BIBentryALTinterwordstretchfactor\fontdimen3\font minus
  \fontdimen4\font\relax}
\providecommand{\BIBforeignlanguage}[2]{{%
\expandafter\ifx\csname l@#1\endcsname\relax
\typeout{** WARNING: IEEEtran.bst: No hyphenation pattern has been}%
\typeout{** loaded for the language `#1'. Using the pattern for}%
\typeout{** the default language instead.}%
\else
\language=\csname l@#1\endcsname
\fi
#2}}
\providecommand{\BIBdecl}{\relax}
\BIBdecl

\bibitem{iPhoneOverloadATT}
\BIBentryALTinterwordspacing
``{Customers Angered as iPhones Overload AT\&T}.'' [Online]. Available:
  \url{http://www.nytimes.com/2009/09/03/technology/companies/03att.html}
\BIBentrySTDinterwordspacing

\bibitem{Chandrasekhar@CMag08}
V.~Chandrasekhar, J.~Andrews, and A.~Gatherer, ``Femtocell networks: a
  survey,'' \emph{Communications Magazine, IEEE}, vol.~46, no.~9, pp. 59 --67,
  september 2008.

\bibitem{Choi@GLOBECOM08}
D.~Choi, P.~Monajemi, S.~Kang, and J.~Villasenor, ``Dealing with loud
  neighbors: The benefits and tradeoffs of adaptive femtocell access,'' in
  \emph{IEEE GLOBECOM}, 30 2008-dec. 4 2008.

\bibitem{Xuejun@ICNP11}
X.~Zhuo, W.~Gao, G.~Cao, and Y.~Dai, ``Win-coupon: An incentive framework for
  3g traffic offloading,'' in \emph{IEEE ICNP}, 2011.

\bibitem{Hua@TMM11}
S.~Hua, Y.~Guo, Y.~Liu, H.~Liu, and S.~Panwar, ``Scalable video multicast in
  hybrid 3g/ad-hoc networks,'' \emph{IEEE Trans. on Multimedia}, vol.~13,
  no.~2, pp. 402--413, 2011.

\bibitem{Hua@Globecom09}
------, ``Sv-bcmcs: Scalable video multicast in hybrid 3g/ad-hoc networks,'' in
  \emph{IEEE GLOBECOM}, 2009, pp. 1--6.

\bibitem{AuctionBook}
V.~Krishna, \emph{{{Auction Theory}}}.\hskip 1em plus 0.5em minus 0.4em\relax
  Academic Press, 2002.

\bibitem{Yanjiao@INFOCOM12}
Y.~Chen, J.~Zhang, Q.~Zhang, and J.~Jia, ``A reverse auction framework for
  access permission transaction to promote hybrid access in femtocell
  network,'' in \emph{IEEE INFOCOM Mini Symposium}, 2012.

\bibitem{Yanjiao@TWC12}
Y.~Chen, J.~Zhang, and Q.~Zhang, ``{Utility-Aware Refunding Framework for
  Hybrid Access Femtocell Network},'' \emph{Wireless Communications, IEEE
  Transactions on}, vol.~11, no.~5, pp. 1688--1697, May 2012.

\bibitem{Dobzinski@JAIR12}
S.~Dobzinski and N.~Nisan, ``{Mechanisms for Multi-Unit Auctions},''
  \emph{Journal of Artificial Intelligence Research}, vol.~37, no.~1, 2010.

\bibitem{Christos@BOOK07}
C.~H. Papadimitriou and K.~Steiglitz, \emph{{Combinatorial Optimization:
  Algorithms and Complexity}}.\hskip 1em plus 0.5em minus 0.4em\relax Dover
  Publications, 1998.

\bibitem{Myerson@JET83}
R.~B. Myerson and M.~A. Satterthwaite, ``Efficient mechanisms for bilateral
  trading,'' \emph{Journal of Economic Theory}, vol. 29(2), pp. 265--281, April
  1983.

\bibitem{Xuejun@TMC13}
X.~Zhuo, W.~Gao, G.~Cao, and S.~Hua, ``An incentive framework for cellular
  traffic offloading,'' \emph{IEEE Trans. on Mobile Computing}, to appear.

\bibitem{WiMAXEva}
\BIBentryALTinterwordspacing
I.~802.16m 08/004r1, ``{802.16m evaluation methodology document},'' March 2008.
  [Online]. Available:
  \url{http://wirelessman.org/tgm/docs/80216m-08_004r1.pdf}
\BIBentrySTDinterwordspacing

\bibitem{FemtoForum}
\BIBentryALTinterwordspacing
FemtoForum, ``{Interference Management in OFDMA Femtocells},'' March 2010.
  [Online]. Available: \url{www.femtoforum.org}
\BIBentrySTDinterwordspacing

\bibitem{Sha@ICC12}
S.~Hua, P.~Liu, and S.~Panwar, ``The urge to merge: When cellular service
  providers pool capacity,'' in \emph{IEEE ICC}, 2012.

\end{thebibliography}

\end{document}